\documentclass[reprint,nofootinbib,amsmath,amssymb,prb,showpacs,citeautoscrip]{revtex4-1}
\usepackage{graphicx}
\usepackage{bm}
\usepackage{amssymb}   
\usepackage{xcolor, colortbl}
\usepackage{textcomp}
\usepackage{natbib}
\usepackage{array}

\definecolor{darkseagreen1}{rgb}{0.82,0.94,0.82}

\begin{document}

 \title{Enhanced piezoelectricity and modified dielectric screening of 2-D group-IV monochalcogenides}

 \author{L\'idia C. Gomes}
 \affiliation{ Centre for Advanced 2D Materials and Graphene Research Centre, National University of Singapore, 6 Science Drive 2, 117546, Singapore }%
 \author{A. Carvalho}
 \affiliation{ Centre for Advanced 2D Materials and Graphene Research Centre, National University of Singapore, 6 Science Drive 2, 117546, Singapore }%
 \author{A. H. Castro Neto}
 \affiliation{ Centre for Advanced 2D Materials and Graphene Research Centre, National University of Singapore, 6 Science Drive 2, 117546, Singapore }%
 \date{\today}

\begin{abstract}
We use first principles calculations to investigate the lattice properties of group-IV monochalcogenides.
These include static dielectric permittivity, elastic and piezoelectric tensors.
For the monolayer, it is found that the static permittivity, besides acquiring a dependence on the interlayer distance,
is comparatively higher than in the 3D system.
In contrast, it is found that elastic properties are little changed by the lower dimensionality.
Poisson ratio relating in-plane deformations are close to zero, and the existence of a negative Poisson
ratio is also predicted for the GeS compound.
Finally, the monolayers shows piezoelectricity, with piezoelectric constants higher than that recently predicted to occur in  
other 2D-systems, as hexagonal BN and transition metal dichalcogenide monolayers. 
\end{abstract}

\pacs{Valid PACS appear here}
\maketitle


\hyphenation{ALPGEN}
\hyphenation{EVTGEN}
\hyphenation{PYTHIA}

\section{Introduction}

Piezoelectric materials are used in very diverse fields of application as sensors, actuators,
electric field generators, and in general in any other applications requiring a conversion between 
electrical and mechanical energy.
Recently, the availability of piezoelectric nanowires and nanobelts inspired 
the design of a class of `nanopiezotronic' devices 
which make use of both their piezoelectric and semiconducting properties.\cite{wang}
These include, for example, nano-generators, field-effect transistors, and piezoelectric diodes.
Lead-free biocompatible piezoelectric materials are also sought for, as components for
biomedical devices.

Piezoelectric crystals can also be found amongst 2D materials.
Some of these, like BN and 2H-stacked transition
metal dichalcogenides, are centrosymmetric in bulk form,
but loose the inversion symmetry if the number of layers is odd.
Exfoliation of W and Mo dichalcogenide monolayers thus produces
2D piezoelectric crystals with strain-induced polarisation change in plane.
This has been confirmed experimentally for MoS$_2$ by 
direct piezoresponse measurements
and electrical characterisation of MoS$_2$ devices under strain\cite{wu-N-514-470}.
For a few transition metal dichalcogenides, the in-plane $d_{11}$ components of the piezoelectric tensor
have been predicted to be
superior to quartz.\cite{jz3012436}

Here, we concentrate on yet another family of layered materials
that are piezoelectric in the monolayer form.
These are the group-IV monocalcogenides\cite{PhysRevB.92.085406}
SnS, GeS, SnSe and GeSe.
Due to the hinge-like structure similar to that of black
phosphorus~\cite{acsnano.5b02742} (Fig.~\ref{structure}), group-IV monocalcogenides are very ductile along the direction perpendicular to 
the zigzags, stretching in that direction when out-of-plane strain is applied.
Since the direction perpendicular to the zigzags is also the main polar direction,
this results in a very large piezoelectric coefficient, exceeding by at least one order of magnitude that of
other known 2D piezoelectrics.

In addition, group-IV monochalcogenides present nearly vanishing or negative Poisson ratio.
This is also a consequence of their anisotropic structure, as was recently reported~\cite{nature-5727} that a monolayer of black phosphorus (phosphorene) also possesses a negative Poisson ratio. 

In this article, we use first principles calculations to predict the lattice response properties of this family of materials,
comprising static dielectric constant, and elastic and piezoelectric constants.

\subsection{Computational details}

We use first-principles calculations to obtain structural properties of monochalcogenides. 
We employ a first-principles approach based on Kohn-Sham density functional theory (KS-DFT)\cite{PhysRev.140.A1133}, as implemented 
in the Vienna ab initio simulation package (VASP)~\cite{Kresse199615, PhysRevB.54.11169}, which was used for calculation of elastic, piezoelectric and static dielectric tensors. The core and valence electrons are treated with the projector-augmented wave (PAW) method~\cite{PhysRevB.50.17953}.

The exchange correlation energy was described by the generalized gradient approximation (GGA) using the PBE\cite{PhysRevLett.77.3865} functional. For all materials, van der Waals interactions are taken into account by the method proposed by Tkatchenko and Scheffler~\cite{PhysRevLett.102.073005} (TS), which presents a charge-density dependence of the dispersion coefficients and damping function. The Kohn-Sham orbitals were expanded in a plane-wave basis with a cutoff energy of 550~eV. The Brillouin-zone (BZ) was sampled using a $\Gamma$-centered 1$\times$20$\times$20 grid for the monolayers, following the scheme 
proposed by Monkhorst-Pack\cite{PhysRevB.13.5188}. The results for bulk were obtained with a 6$\times$16$\times$16 grid. Structural optimization has been performed with a very stringent tolerance of 0.001~eV/\AA{}. 

For comparison, we also performed calculations with the {\sc Quantum ESPRESSO} code,\cite{Giannozzi2009} in that case using Troullier-Martins pseudopotentials\cite{PhysRevB.43.1993}. 
There was good agreement between the two codes whenever direct comparison could be established.
All results shown were obtained with VASP, with exception of the Poisson ratio calculations,
which were performed with {\sc Quantum ESPRESSO} to take advantage of the implementation of geometry optimisation constraints.
 
The supercells are periodic in the monolayer plane and large vacuum regions ($>$ 17~\AA{}) are included to impose 
periodic boundary conditions in the perpendicular direction. Convergence tests with greater vacuum thickness were 
performed, and the values used are enough to avoid spurious interaction between neighboring images.
 
\section{Results}

\subsection{Structure}

\begin{figure}[!htb]
    \centerline{
    \includegraphics[scale=0.32]{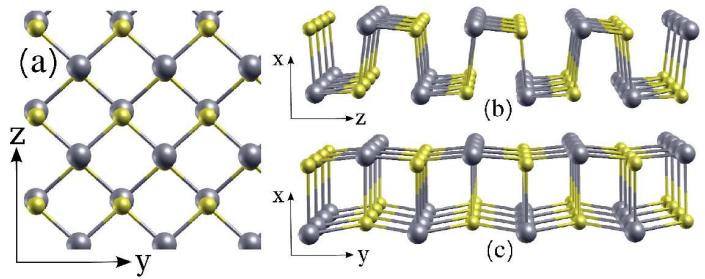}}
   \caption{\small (Color online) Structure of group-IV monochalcogenides. (a) Top-view and (b, c) side views  of the monolayer unitcell, which are repeated along the $x$ direction to obtain the bulk. The layers sit on the $y$-$z$ plane, with the $y$ axis parallel to the puckering direction.}
   \label{structure}
 \end{figure}

Bulk group-IV monochalcogenides SnS, SnSe, GeS and GeSe have an orthorhombic structure with eight atoms per unit cell, four of each species. They belong to the space group $Pnma$.
This structure is also known as the $\alpha$ phase of SnS, a naturally occurring mineral~\cite{PhysRevB.70.235114}. 
The waved structure adopted by these materials is similar to that of black phosphorus, with which these compounds are isoelectronic.
All atoms are three-fold coordinated and tetrahedral coordination results from the repulsion of the lone pairs. 
In the monolayer, the translational symmetry along the $x$ direction is lost, and with it the inversion symmetry, 
and the resulting structure belongs to the $Pmn2_{1}$ space group. In this work, the axes system is chosen as the layers sit on the $y$-$z$ plane, with the $y$ axis parallel to the puckering direction, as shown in Fig.~\ref{structure}. The layers are stacked together along the $x$ axis to form the bulk.

The calculated lattice parameters of these materials, for both bulk and monolayer, have been discussed in previous theoretical studies,\cite{jap.113.23-10.1063, apl.100.3-10.1063, Alptekin2010870}, including our previous work at the GGA level~\cite{PhysRevB.92.085406}. 
Aware of the importance of van der Waals (vdW) interactions in layered materials, in this work we include these effect in order to obtain even more accurate structural properties of these materials. Table~\ref{vectors} presents the optimized lattice parameters for pure GGA and including vdW effects. Available experimental data for bulk SnS and SnSe are also included. 

As should be expected, the most noticeable effects of vdW interactions in the structural parameters are on the lattice vector $\mathbf{a}$, the one perpendicular to the plane of the layers in bulk. The inclusion of vdW forces results in values for $\mathbf{a}$ very close to the experimental data for bulk SnS and SnSe~\cite{wiedemeier, LEhm, PhysRevB.58.1896}, with better results for the Tkatchenko-Scheffler method. The difference between PBE and vdW for monolayers is minor.

\begin{table}[ht]
\centering
\renewcommand{\arraystretch}{1.6}
\begin{tabular}{lclccccccc}
        \hline \hline
    &&&\phantom{a}& \multicolumn{2}{c}{Monolayer}&\phantom{abc}&\multicolumn{3}{c}{Bulk}  \\ 
         \cline{5-6} \cline{8-10}
    &&&&$\mathbf{b}$&$\mathbf{c}$&&$\mathbf{a}$&$\mathbf{b}$&$\mathbf{c}$ \\ \hline
SnS  && PBE   &&   4.07     &  4.24      &&    11.37   &    4.02    &   4.35     \\
     &&vdW-TS &&   4.08     &  4.25      &&    11.12   &    4.00    &   4.27     \\ 
     && Expt. &&   -        &    -       &&    11.20   &    3.98    &   4.33     \\ \hline

SnSe && PBE   &&   4.30     &  4.36      &&    11.81   &    4.22    &   4.47     \\
     &&vdW-TS &&   4.27     &  4.37      &&    11.58   &    4.20    &   4.47     \\
     && Expt. &&   -        &    -       &&    11.50   &    4.15    &   4.44     \\ \hline

GeS  && PBE   &&   3.68     &  4.40      &&    10.81   &    3.68    &   4.40     \\
     &&vdW-TS &&   3.73     &  4.30      &&    10.41   &    3.67    &   4.34     \\ \hline

GeSe && PBE   &&   3.99     &  4.26      &&    11.31   &    3.91    &   4.45     \\
     &&vdW-TS &&   4.00     &  4.22      &&    10.93   &    3.90    &   4.39     \\
\hline \hline
\end{tabular}
\caption{\small Optimized lattice parameters (in \AA{}) for $\alpha$ phase of SnS, SnSe, GeS, GeSe. We present results for PBE-GGA and including vdW by Tkatchenko-Scheffler (TS) method. Experimental data for bulk SnS~\cite{wiedemeier, LEhm} and SnSe~\cite{PhysRevB.58.1896} are also presented.}
\label{vectors}
\end{table}

\subsection{Static Dielectric Tensor}

The static dielectric tensor $\epsilon$ is calculated from the force constant ($\bf K$) and  
Born dynamical effective charges (BEC) ($\bf Z$) tensors:\cite{PhysRevB.72.035105}

\begin{equation}
\epsilon_{ij}=\epsilon^{\infty}_{ij}+V_0^{-1}Z_{mi}(K^{-1})_{mn}Z_{nj}.\label{epscalc}
\end{equation}

The tensors $\bf K$ and $\bf Z$ are second derivative response functions and are calculated using density functional perturbation theory.

The electronic contribution $\epsilon^{\infty}_{ij}$, or ion-clamped static dielectric tensor,
has been given in Ref.~\onlinecite{PhysRevB.92.085406}. 
Here, we will concentrate on the lattice contribution,
which for static electric fields, exceeds the electronic contribution.

For orthorhombic materials, the dielectric permittivity tensor has the form
\begin{equation}
\epsilon_{ij} =  \left[
\begin{array}{c c c }
   \epsilon_1   &      0            &      0     \\ 
      0         &      \epsilon_2   &      0     \\ 
      0         &      0            &      \epsilon_3     \\ 
\end{array}
          \right],
\end{equation}
where we have contracted a two-index into a one-index notation.
The values calculated for bulk are given in Tab.~\ref{dielec}.
While for the tin chalcogenides the calculated values are in excellent agreement with 
estimates obtained from comparing the LO-TO splitting of the vibrational bands
with the dielectric constants obtained from reflectivity data,
for the germanium chalcogenides there is some discrepancy. 
This can be due to many factors, including crystal quality.
In particular, the dielectric response along the $\bf x$ axis is difficult to measure, explaining the range of variation of the measured values.

For two-dimensional materials, the dielectric constant is not well defined, depending on the interlayer distance $L$ as~\cite{PhysRevB.84.085406, PhysRevB.88.045318, jap.34-7.1853}

\begin{equation}
\epsilon_{i}=\delta_{i}+\frac{4\pi\chi_{i}^{\rm 2D}}{L},
\end{equation}
where the 2D polarizability $\chi^{\rm 2D}$ (which includes both ionic and electronic contributions) is a constant and $\alpha,\beta=2,3$, in accordance 
with our convention of the layers in the $y-z$ plane. The 1/$L$ dependence on the interlayer spacing of the electronic contribution 
to the susceptibility in the 2D systems has been verified in previous works.~\cite{PhysRevB.84.085406, PhysRevB.88.045318, jap.34-7.1853}. In this work, 
we verified that the same behaviour applies to the ionic contribution. In practice, this can be calculated using the same method as bulk (Eq.~\ref{epscalc}) but replacing the volume by the area. The ionic contribution 
to $\chi^{\rm 2D}$ for the monolayers is given in Tab.~\ref{dielec}. 

\begin{table*}[htb!]
\centering
\renewcommand{\arraystretch}{1.6}
\begin{tabular}{lclcccccccccccccccc}
     \hline \hline
     &&&& \multicolumn{5}{c}{Monolayer}  &  &\multicolumn{9}{c}{Bulk}  \\  
     \cline{5-9}   \cline{11-19}
     &&&&&&&&&&&&&&\multicolumn{5}{c}{Expt.}     \\  
     \cline{15-19}
 &\phantom{ab}&&\phantom{ab}&$\chi_{2}^{\rm 2D}$&$\chi_{3}^{\rm2D}$&\phantom{abc}&$\epsilon_{2}$&$\epsilon_{3}$&\phantom{abc}&$\epsilon_{1}$&$\epsilon_{2}$&$\epsilon_{3}$&\phantom{abc}&$\epsilon_{1}$&$\epsilon_{2}$
 &$\epsilon_{3}$&\phantom{ab}& \\ \hline
SnS  && PBE   &&  35.8    &   21.5   &&     78.0      &     50.1      && 14.9   & 31.7  & 18.3  &&16$\pm$6 &32$\pm$7 & 18$\pm$6 && Ref.~(a)  \\
                                                                                                
     &&vdW-TS &&  27.3    &   17.2   &&     61.7      &     39.0      && 27.5   & 22.6  & 13.8  &&  18.4   &   34.5  &   20.7   && Ref.~(d)  \\ \hline
                                                                                                                           
SnSe && PBE   &&  73.6    &   26.4   &&    156.5      &     56.0      && 21.3   & 39.1  &12.1   &&26$\pm$7 &45$\pm$8 & 32$\pm$7 && Ref.~(a)  \\
                                                                                                
     &&vdW-TS &&  76.4    &   37.7   &&    165.7      &     82.0      && 25.7   & 32.0  &15.9   &&  17.5   &   32.5  &   18.5   && Ref.~(d)  \\ \hline
                                                                                                                                 
GeS  && PBE   &&   5.3    &    3.1   &&     12.4      &      7.4      &&  6.8   & 11.8  & 7.6   &&  20.0   &   17.5  &   10.3   && Ref.~(b)  \\

     &&vdW-TS &&   7.1    &    3.6   &&     16.1      &      8.6      && 16.7   & 12.9  & 8.5   &&  10.6   &   14.0  &   10.6   && Ref.~(d)  \\ \hline
                                                                                                                                 
GeSe && PBE   &&  24.0    &    7.7   &&     53.3      &     17.1      && 6.9    & 12.3  &  5.7  &&  11.3   &    8.5  &    3.2   && Ref.~(c)  \\

     &&vdW-TS &&  28.6    &   10.1   &&     65.7      &     24.3      && 14.3   & 15.7  &  7.7  &&   9.7   &   12.5  &    9.6   && Ref.~(d)  \\
 \hline \hline
\end{tabular}
\caption{\small Ionic contributions to the 2D polarizability $\chi_{2}^{\rm 2D}$ and $\chi_{3}^{\rm 2D}$, macroscopic static dielectric tensor components $\epsilon_{2}$ and $\epsilon_{3}$ for $y$ and $z$ directions of the monolayers, and $\epsilon_{1}$, $\epsilon_{2}$ and $\epsilon_{3}$ for $x$, $y$ and $z$ directions of bulk. Results for PBE-GGA and taking into account van der Walls interactions are included. Our calculations for the bulk can be compared to experimental values from the previous works: (a)~Ref.~\onlinecite{PhysRevB.15.2177}, (b)~Ref.~\onlinecite{PhysRevB.13.2489}, (c)~Ref.~\onlinecite{ssc.18.11.1509}, (d)~Ref.~\onlinecite{PhysRevB.47.16222}.} 
\label{dielec}
\end{table*}

For comparison, if the interlayer distance $L$ in the monolayer is taken to be the same as in bulk, the vdW-TS calculations for SnS results in static ionic contributions to the permittivity with values 61 and 23, respectively, along the $y$ direction, and 39 and 14, respectively, along the $z$ direction, showing an increase of $\sim$~2.5 times when comparing the 2D and 3D systems. 
For SnSe and GeSe the ionic contribution to $\epsilon$ in monolayer reaches values three to four times higher than in their respective bulk forms. 
In the other hand, GeS presents just slightly higher values for $\epsilon$ from bulk to monolayer.
The increase in the screening for SnS, SnSe and GeSe can be explained by two factors: (i) increase in Born effective charges (BEC) and (ii) softening of polar modes due to the absence of interlayer binding. The increase in the BEC can be as higher as 28\% for SnS, SnSe and GeSe, while for GeS $\bf Z$ differs by less than 3\%. 
To have a better understanding of the contribution of lattice vibrational mode softening to the increase of the static dielectric constant, we consider the contribution of different modes to the components of the static dielectric tensor, by writing $\epsilon_{ij}$ as in Ref.~\onlinecite{PhysRevB.55.10355, PhysRevB.63.104305}:

\begin{equation}
\epsilon_{ij} = \epsilon^{\infty}_{ij} + \dfrac{4\pi}{V_{0}}\sum_{m} \dfrac{S_{m,ij}}{\omega^{2}_{m}}
\label{eps-ab}
\end{equation}
where $V_{0}$ is the volume of the cell. $S_{m,ij}$ is called the mode-oscillator strength 
of the $\omega_{m}$ mode along the $i,j$(=$x,y,z$) directions of the material and it is related to the eigendisplacements 
$U_{m}(\kappa i)$ and Born effective charge tensors $Z^{*}_{\kappa,ii'}$ by:

\begin{equation}
S_{m,ij} = \left(\sum_{\kappa i'} Z^{*}_{\kappa,ii'} U^{*}_{m{\bf q}=0}(\kappa i')\right)\left(\sum_{\kappa' j'} Z^{*}_{\kappa',jj'} U_{m{\bf q}=0}(\kappa' j')\right).
\label{sm-ab}
\end{equation}
$\kappa$ is the index for the ions in the primitive cell. The ionic contribution to $\epsilon$ is then given by the second term in the right side of Eq.~\ref{eps-ab}. The expression for $S_{m,\alpha \beta}$ shows that and increase in the BEC increases the contribution of the ionic part of 
$\epsilon$. However, the dominant contribution comes from the $\omega_{m}$ modes. In Tab.~\ref{omega} we present the frequency of the modes with major contribution to the static dielectric constant for monolayer and bulk monochalcogenides, along the in-plane directions. For all materials, just one mode contributes significantly to the $yy$ component of $\epsilon$, while there are two modes which contributes almost equally to the $zz$ components in bulk SnS and monolayer and bulk SnSe and GeSe. A significant softening of the contributing modes occurs when going from 3D to 2D forms, probably due to the lack of interaction with upper and bottom layers, observed in bulk. SnSe presents the largest softening in the contributing modes, with $\omega_{m}^{3D}$/$\omega_{m}^{2D}$ = 1.9 along the $y$ direction up to $\omega_{m}^{3D}$/$\omega_{m}^{2D}$ = 2.5 along $z$. GeS, however, presents the smallest softening of the modes, where we have $\omega_{m}^{3D}$/$\omega_{m}^{2D}$ = 1.1 and 1.5 for the contribution to the $yy$ and $zz$ components of $\epsilon$.

\begin{table}[htb!]
\centering
\renewcommand{\arraystretch}{1.6}
\begin{tabular}{lccccc}
     \hline \hline
     &\multicolumn{2}{c}{yy}&\phantom{abc}&\multicolumn{2}{c}{zz} \\ 
     \cline{2-3} \cline{5-6}
     &  Monolayer &  Bulk   &&  Monolayer & Bulk        \\  \hline
SnS  &   104      &  158    &&    62      & 102 / 189   \\
                                             
SnSe &   53       &  102    &&  56 / 78   & 132 / 86    \\
                                             
GeS  &  180       &  207    &&    77      &  116        \\
                                             
GeSe &   94       &  145    && 137 / 76   & 184 / 87    \\
 \hline \hline
\end{tabular}
\caption{\small Modes (cm$^{-1}$) at {\bf q}=0 with largest contribution to the static dielectric tensor as defined in Eq. \ref{eps-ab}. The values presented in this table are obtained considering vdW interactions and do not take into account LO-TO splitting effects.}
                                
\label{omega}
\end{table}

\subsection{Elastic Constants and Poisson ratio}

\paragraph{Elastic constants:}

The elastic constant (stiffness) tensor is defined in the linear regime
by Hooke's law:

\begin{equation}
 \sigma_{i} = C_{ij}\varepsilon_{j},
 \label{elastic-c}
\end{equation}
where $\sigma$ and $\varepsilon$ are the stress and strain tensors, respectively, $C_{ij}$ are 
the elastic constants. 
In this equation, and the following, there is an implicit sum over repeated indexes.
Numerically, the elastic constants are easily determined using the equation of state.
The change in the total energy $E(V_0)$ of a system due to an external strain ($\varepsilon$) is given by
\begin{equation}
E(V_0,\varepsilon) = E(V_0)+ \frac{V_{0}}{2}C_{ij}\varepsilon_{i}\varepsilon_{j},
 \label{etot-cij}
\end{equation}
where $V_{0}$ is the volume of the unstrained material.
Here, the electric field term vanishes because of the periodic boundary conditions ($d\vec{E}=0$).
For the monolayer system, the area ($A_{0}$) should figure instead.
However, to facilitate the comparison with 3D systems and previous calculations,\cite{wei-arxiv1403.7882} we use an effective thickness $d_0$. The most obvious choice is to take $d_0$ to be the distance between the layers in bulk, as is common practice in graphene~\cite{jpcm-21-33}.

Symmetry imposes restrictions on the number of non-zero components $C_{ij}$. 
Both for bulk and monolayer there are nine independent nonvanishing 
elements $C_{ij}$ (in Voigt notation):
\begin{equation}
C_{ij} =  \left[
\begin{array}{c c c c c c c}
   C_{11}     &   C_{12}    &  C_{13}   &      0     &      0     &      0     \\
   C_{12}     &   C_{22}    &  C_{23}   &      0     &      0     &      0     \\        
   C_{13}     &   C_{23}    &  C_{33}   &      0     &      0     &      0     \\
      0       &     0       &     0     &   C_{44}   &      0     &      0     \\ 
      0       &     0       &     0     &      0     &   C_{55}   &      0     \\ 
      0       &     0       &     0     &      0     &      0     &  C_{66}    \\ 
\end{array}
          \right] 
\end{equation}

However, we will consider that the monolayers are ideal 2D systems, with only in-plane stress.
This excludes the elastic constants coupling $\varepsilon_{i}$ to $\sigma_{6}$ 
and $\sigma_{5}$, (this is equivalent to $\sigma_{12}$ and $\sigma_{13}$).
Since in the absence of an external torque the $\sigma$ tensor is symmetrical, the only relevant elements are
(in Voigt's notation) $C_{22}$, $C_{33}$, $C_{23}$ and $C_{44}$.

The elastic constants for the monolayer were obtained from Eq.~\ref{elastic-c},
 applying finite distortions to the supercell. 
The stress components $\sigma_{2i}$ and $\sigma_{3i}$ are calculated using 
the effective thickness $d_0$ to define the lateral area of the layers. 
We have calculated $C_{ij}$ both in clamped-ion  and relaxed-ion conditions,
by fixing the atomic coordinates or allowing them to relax, respectively,
for each distortion of the lattice vectors.
The results for bulk and monolayer group-IV monochalcogenides are presented in Table~\ref{Cij}, for calculations with and without vdW effects. 
Using the definition of $d_{0}$, the elastic constants are of the same order of magnitude for monolayer and bulk. In most cases though, they are slightly higher for the bulk for both relaxed and clamped-ion coefficients, indicating that the monolayers are less stiff for in-plane deformations. The main effects of vdW interactions in bulk are observed in the elastic constant components related to the $x$ direction of the materials (those with index $1j$, j=$1,2,3$ as well as $55$ and $66$). The $C_{11}$ constants are the most affected by the vdW forces, as expected, and differences up to 50\% are observed when compared to the PBE-GGA results.

\begin{figure}[!htb]
    \centerline{
     \includegraphics[scale=0.22]{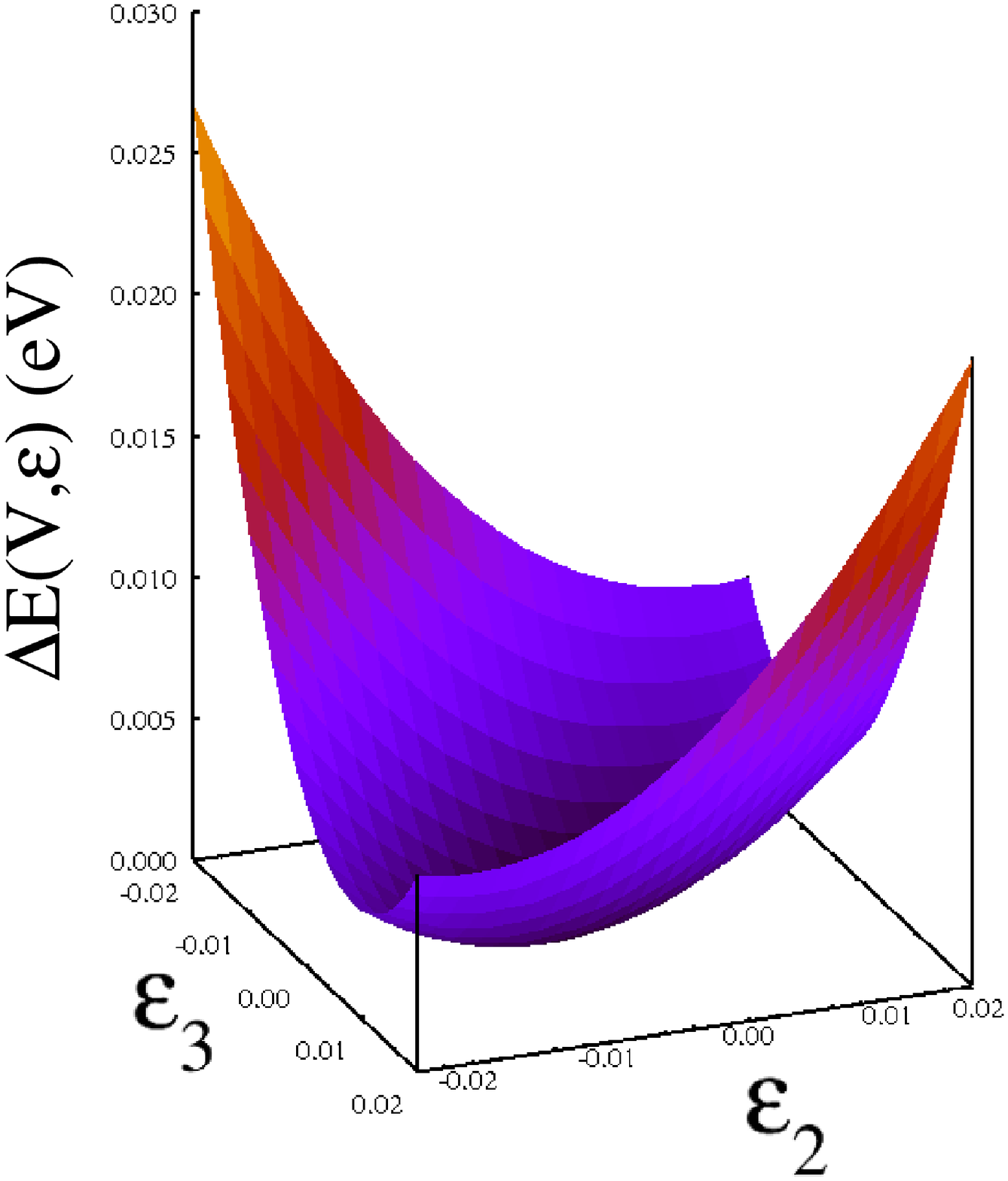}
     \includegraphics[scale=0.22]{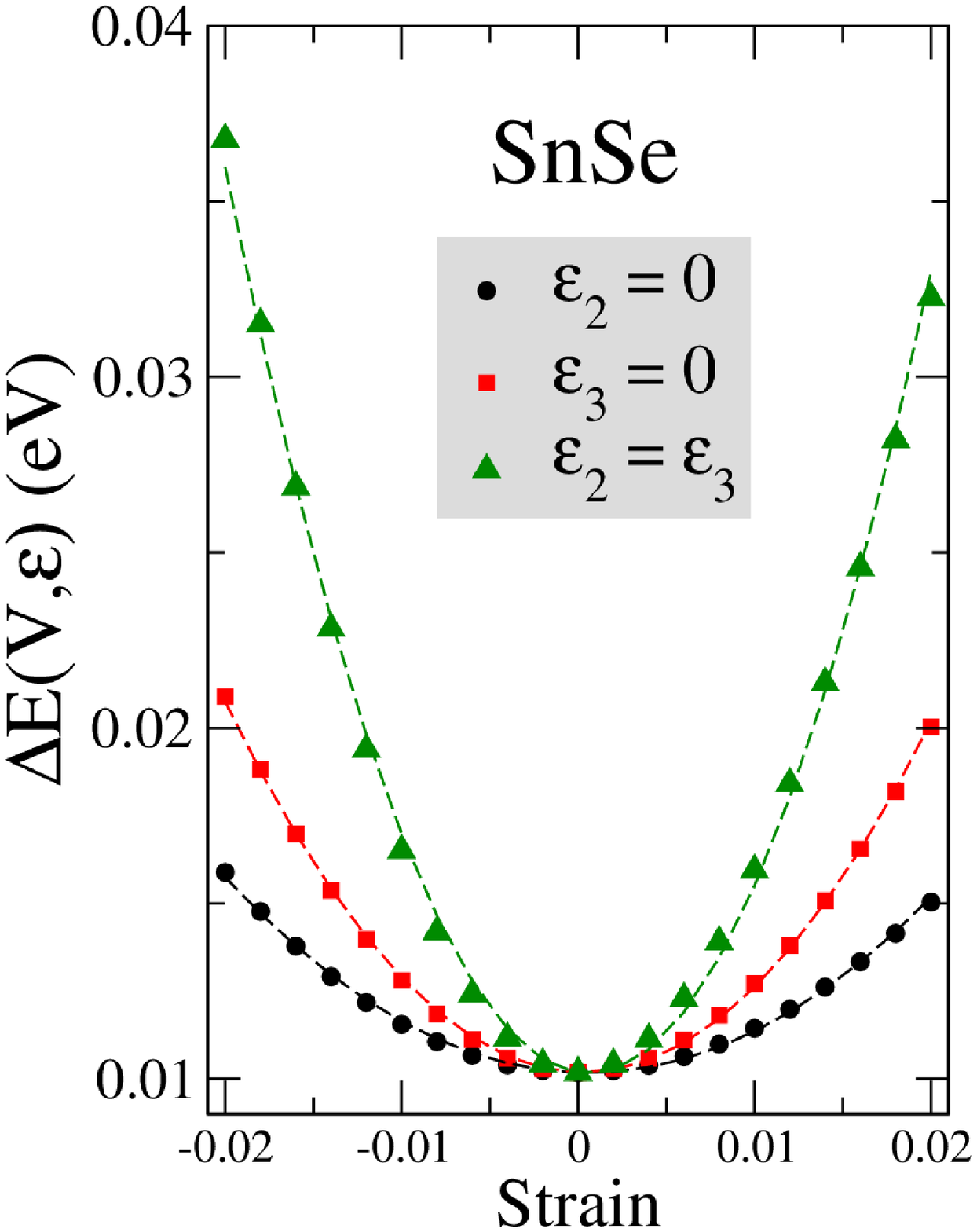}}
   \caption{(Color online) \small The three-dimensional surface plot of total energy (in eV) versus uniaxial strain along $y$ and $z$ directions of a SnSe monolayer. Projections on the $\varepsilon_{2}$~=~0, $\varepsilon_{3}$~=~0 and $\varepsilon_{2}$~=~$\varepsilon_{3}$ planes show the quadratic dependence of the energy of the system to the applied strain, from where the elastic constants can also be calculated.}
   \label{Etot-sy-sz}
 \end{figure}

Figure~\ref{Etot-sy-sz} shows the total energy versus strain along the $y$ and $z$ directions 
for a SnSe monolayer. The quadratic dependence of energy to the applied strain and the anisotropy 
between the perpendicular in-plane directions of the layer is given by
\begin{equation} 
\Delta E(V,\varepsilon) = \frac{V_0}{2}\left(C_{22}\varepsilon^2_{2} + C_{33}\varepsilon^2_{3}\right) + V_0 C_{23}\varepsilon_{2}\varepsilon_{3} 
\end{equation}
where $\Delta E(V,\varepsilon) = E(V,\varepsilon) - E(V_0,\varepsilon=0)$ and $C_{32}~=~C_{23}$.
The results for SnS, GeS and GeSe are very similar, differing only in the ratio between elastic constants.

It is also noteworthy that 
$C_{22}$ (the elastic constant for the zig-zag direction) is much lower for group-IV monochalcogenides than for black-phosphorus (18.6 $\times10^{10}$ N/m$^2$),
in contrast with other elastic constant elements which are little changed~\cite{PhysRevB.86.035105}.

\begin{table*}[htb!]
\centering
\renewcommand{\arraystretch}{1.5}
\begin{tabular}{m{1.0cm} m{0.3cm} m{1.4cm} m{0.3cm} m{0.8cm} m{1.3cm} m{1.3cm} m{1.3cm} m{1.3cm} m{1.3cm} m{1.3cm} m{1.3cm}  m{1.3cm} m{0.9cm}}
     \hline  \hline    
     &&&&&       \multicolumn{9}{c}{Monolayer}                                                              \\ \hline
    &&     &&     &  $C_{11}$ &   $C_{22}$   & $C_{33}$  &  $C_{12}$ & $C_{13}$    &    $C_{23}$  &   $C_{44}$ &  $C_{55}$   & $C_{66}$     \\ \cline{6-14}
SnS && PBE && C-i &     -     &     9.59     &   8.81    &     -     &     -       &      5.68    &    5.50    &     -       &      -       \\
    &&     && R-i &     -     &     7.59     &   3.67    &     -     &     -       &      3.19    &    3.44    &     -       &      -       \\ 
 \rowcolor{darkseagreen1}
 && vdW-TS && C-i &     -     &     9.65     &   9.29    &     -     &     -       &      5.70    &    5.25    &     -       &      -       \\
 \rowcolor{darkseagreen1}
    &&     && R-i &     -     &     7.72     &   3.85    &     -     &     -       &      2.68    &    2.81    &     -       &      -       \\ \hline
                                                                                                                                            
SnSe&& PBE && C-i &     -     &     7.90     &   7.26    &     -     &     -       &    4.54      &    4.81    &     -       &      -       \\              
    &&     && R-i &     -     &     6.92     &   3.32    &     -     &     -       &    2.77      &    2.32    &     -       &      -       \\ 
 \rowcolor{darkseagreen1}
 && vdW-TS && C-i &     -     &     5.68     &   5.34    &     -     &     -       &     2.96     &    3.28    &     -       &      -       \\
 \rowcolor{darkseagreen1}
    &&     && R-i &     -     &     5.04     &   2.89    &     -     &     -       &     1.83     &    1.80    &     -       &      -       \\ \hline
                                                                                                                                           
GeS && PBE && C-i &     -     &     9.51     &   8.92    &     -     &     -       &     6.08     &    5.41    &     -       &      -       \\
    &&     && R-i &     -     &     8.48     &   2.82    &     -     &     -       &     4.00     &    3.44    &     -       &      -       \\
 \rowcolor{darkseagreen1}
 && vdW-TS && C-i &     -     &     9.41     &   9.25    &     -     &     -       &     6.17     &    5.52    &     -       &      -       \\
 \rowcolor{darkseagreen1}
    &&     && R-i &     -     &     8.26     &   3.02    &     -     &     -       &     3.93     &    3.49    &     -       &      -       \\ \hline
                                                                                                                                             
GeSe&& PBE && C-i &     -     &     9.78     &   8.46    &     -     &     -       &     5.34     &    6.00    &     -       &      -       \\
    &&     && R-i &     -     &     8.87     &   3.59    &     -     &     -       &     3.44     &    4.10    &     -       &      -       \\ 
 \rowcolor{darkseagreen1}
 && vdW-TS && C-i &     -     &    10.19     &   9.37    &     -     &     -       &     5.42     &    6.04    &     -       &      -       \\
 \rowcolor{darkseagreen1}
    &&     && R-i &     -     &     9.19     &   4.41    &     -     &     -       &     3.40     &    3.81    &     -       &      -       \\ \hline
    &&&&&      \multicolumn{9}{c}{Bulk}                                                                                                     \\ \hline     
    &&     &&     &  $C_{11}$ &  $C_{22}$   & $C_{33}$  &  $C_{12}$  &   $C_{13}$  &   $C_{23}$  &   $C_{44}$  &   $C_{55}$  &   $C_{66}$   \\ \cline{6-14}
SnS && PBE && C-i &   12.12   &   9.53      &   8.31    &  2.45      &   2.56      &     5.51    &     5.29    &    2.56    &   2.43       \\
    &&     && R-i &    4.85   &   7.99      &   3.42    &  1.31      &   1.80      &     3.16    &     3.47    &    1.95    &   2.00       \\ 
 \rowcolor{darkseagreen1}
 && vdW-TS && C-i &   12.15   &   9.68      &   8.79    &  2.64     &    3.07     &     5.59     &    5.37    &    3.15     &   2.74       \\
 \rowcolor{darkseagreen1}
    &&     && R-i &   6.59    &   7.70       &  3.84     &  1.86     &   2.66      &    2.80      &   3.24     &    2.87     &  2.50       \\ \hline
                                                                                                                                         
SnSe&& PBE && C-i &   10.99   &   8.39      &   7.61    &  1.76      &   2.03      &     5.09    &     5.20     &    1.93    &   1.77       \\
    &&     && R-i &    4.35   &   6.93      &   3.48    &  0.74      &   1.23      &     2.99    &     3.32     &    1.37    &   1.07       \\
 \rowcolor{darkseagreen1}
 &&vdW-TS  && C-i &   11.72   &   9.30      &   8.29    &  1.87      &   2.44      &     5.23    &     5.40     &    2.44    &   1.97       \\
 \rowcolor{darkseagreen1}
    &&     && R-i &   6.09    &   7.49      &   4.27    &  0.99      &   2.12      &     2.86    &     3.29     &    2.09    &   1.24       \\ \hline
                                                                                                                                         
GeS && PBE && C-i &   12.40   &  10.47      &   8.83    &  2.79      &  2.67       &     6.00    &     5.75     &    3.18    &   2.93       \\
           && R-i &    3.71   &  8.96       &   3.02    &  0.97      &  0.72       &     3.31    &     3.65     &    1.69    &   1.73       \\     
 \rowcolor{darkseagreen1}
 &&vdW-TS  && C-i &   12.84   &  10.93       &   9.48    & 2.72      &   2.91      &     5.94     &   6.13     &   3.61      &  3.35        \\
 \rowcolor{darkseagreen1}
    &&     && R-i &    5.31   &  8.94        &  3.41     & 1.11      &   1.44      &     2.81     &    3.76    &   2.90      &  2.51        \\ \hline
                                                                                                                                         
GeSe&& PBE && C-i &   12.14   &   10.47     &  8.59     &  1.98      &  2.05       &     6.45    &     6.52     &    2.43    &   2.42       \\
    &&     && R-i &    3.20   &   8.72      &  3.75     &  0.29      &  0.82       &     3.91    &     3.93     &    1.15    &   0.96       \\ 
 \rowcolor{darkseagreen1}
 &&vdW-TS  && C-i &   12.53   &  10.02       &  8.63     &  1.73     &  2.22       &    5.64      &   5.67     &  2.92       &   2.44       \\
 \rowcolor{darkseagreen1}
    &&     && R-i &   4.16    &  8.06        &  3.85     & 0.26      &  1.38       &    3.05      &   3.50     &   2.08      &   0.88       \\ \hline

\hline \hline 
\end{tabular}
\caption{\small Calculated clamped-ion (C-i) and relaxed-ion (R-i) components of the elastic tensor $C_{ij}$ for bulk and monolayer.
The elastic constants for the monolayer assume an effective layer thickness $d_{0}$ = $a/2$ 
to allow direct comparison with bulk. All values are given in 10$^{10}$~N/m$^{2}$. }
\label{Cij}
\end{table*}

\paragraph{Young modulus and Poisson ratio}

\begin{figure}[!htb]
    \centerline{
     \includegraphics[scale=0.35]{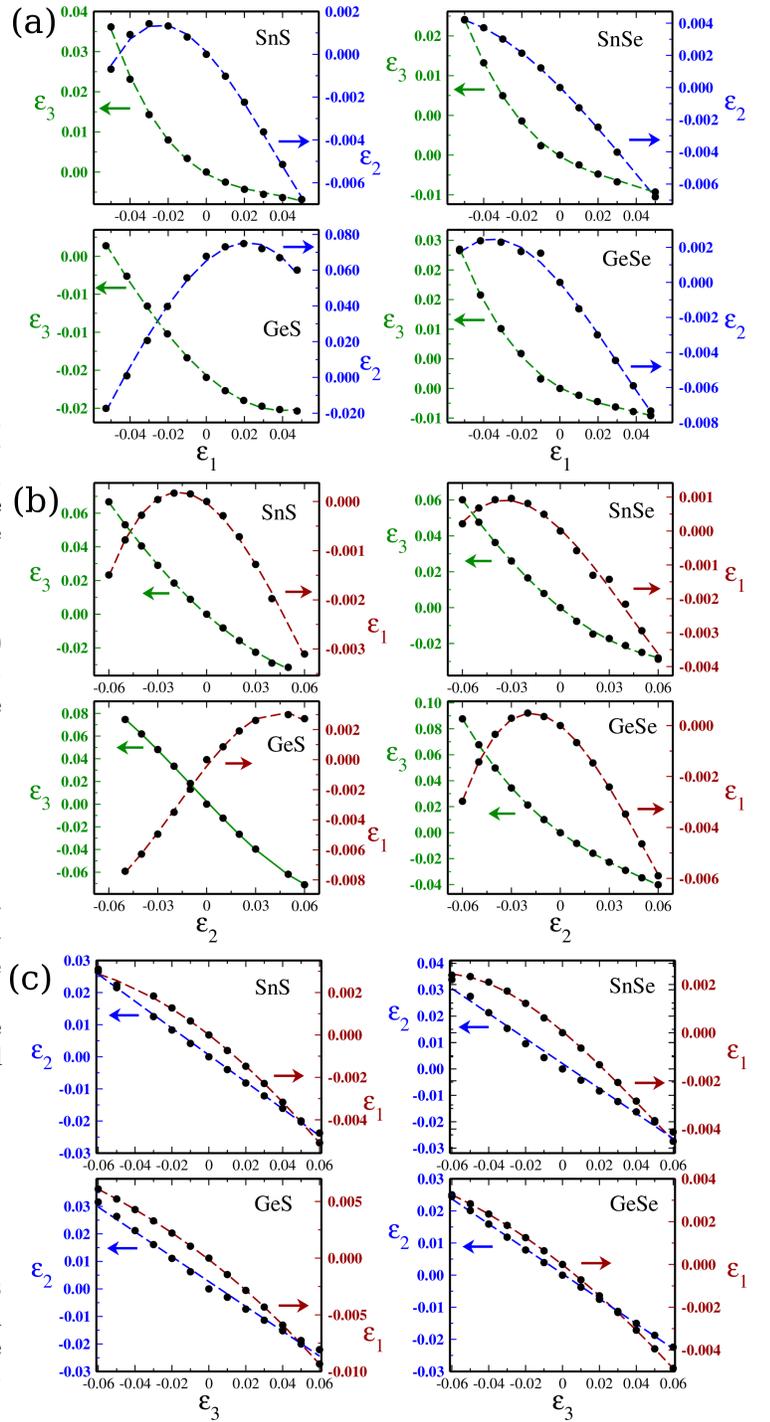}}
   \caption{\small (Color online) Response of monolayer monochalcogenides under uniaxial stress along 
                   (a)~$\hat{x}$, (b)~$\hat{y}$ and (c)~$\hat{z}$ directions of the layers. 
                   Strain components for the directions perpendicular to the external force 
                   are plotted as a function of the strain component along the direction of 
                   the external force.}
   \label{Poisson}
 \end{figure}

The Young modulus and Poisson ratio are derived mechanical properties
that can give direct information on how a system behaves under uniaxial 
stress ie. when $\sigma_{i}\neq0$ and $\sigma_{j}=0$ for all $j\neq i$.
Here, we calculate these moduli to highlight the role of anisotropy on the mechanical 
response of group-IV monochalcogenides.

We define the Young modulus as
\begin{equation}
 Y^i =  \frac{\partial \sigma_i}{\partial \varepsilon_i}.
 \label{Y-eq}
\end{equation}

Since the materials in consideration here are orthorhombic, and the choice of principal direction is arbitrary,
$i$ can be any of the principal directions of the crystal.
Similarly, we define multiple Poisson ratios $\nu_{ij}$,
corresponding to the negative ratio of the strain response 
at a $i$ direction to the strain applied along a transversal $j$ direction:

\begin{equation}
 \nu_{ij} = - \frac{\partial \varepsilon_i}{\partial \varepsilon_j}.
 \label{Poisson-eq}
\end{equation}
for $i\neq j$.

In order to calculate Poisson ratio and Young modulus of monolayer monochalcogenides, according to Eq.~\ref{Y-eq} and \ref{Poisson-eq}, strains from -6\% to 6\% were applied in the in-plane and out-of-plane directions of the layers. As the elastic constants of single-layers are not considerable modified by introduction of vdW effects (Tab.~\ref{Cij}), we include only the results for calculations performed with the PBE-GGA functional.

Most of the materials have a positive Poisson ratio, which means that when a 
compressive strain is applied along one direction, the others expand. Conversely, if the material 
is stretched along one direction, it will compress along the perpendicular directions. 
This due to the materials' tendency  to conserve their volume.
However, since group-IV monochalcogenides are very anisotropic, the values 
for the Poisson ratio vary greatly for different combination of directions $i,j$.

This is apparent in Figure~\ref{Poisson}, which illustrates the response of the monolayers under uniaxial stress.
Firstly, it is clear that the linear region where the Young modulus and Poisson ratio given by
Eqs.~\ref{Y-eq} and ~\ref{Poisson-eq} are constant
is quite narrow, in some cases less than 2\% strain.

In particular, if $i=1$ or $j=1$, the Poisson ratio is very close to zero,
indicating that distortion along the in-plane directions and the direction perpendicular to the plane are
practically decoupled.
A very interesting exception is found in GeS: this is the only material in group-IV monochalcogenides that shows a negative linear Poisson ratio in the out of plane $x$ direction. The calculated value of $\nu_{12}$~=~$-$0.137 for GeS is about five times larger than the value calculated for phosphorene in Ref.~\onlinecite{Jiang2014}, which report $\nu_{12}$~=~$-$0.027 for this material. 

\begin{figure}[!htb]
    \centerline{
     \includegraphics[scale=0.32]{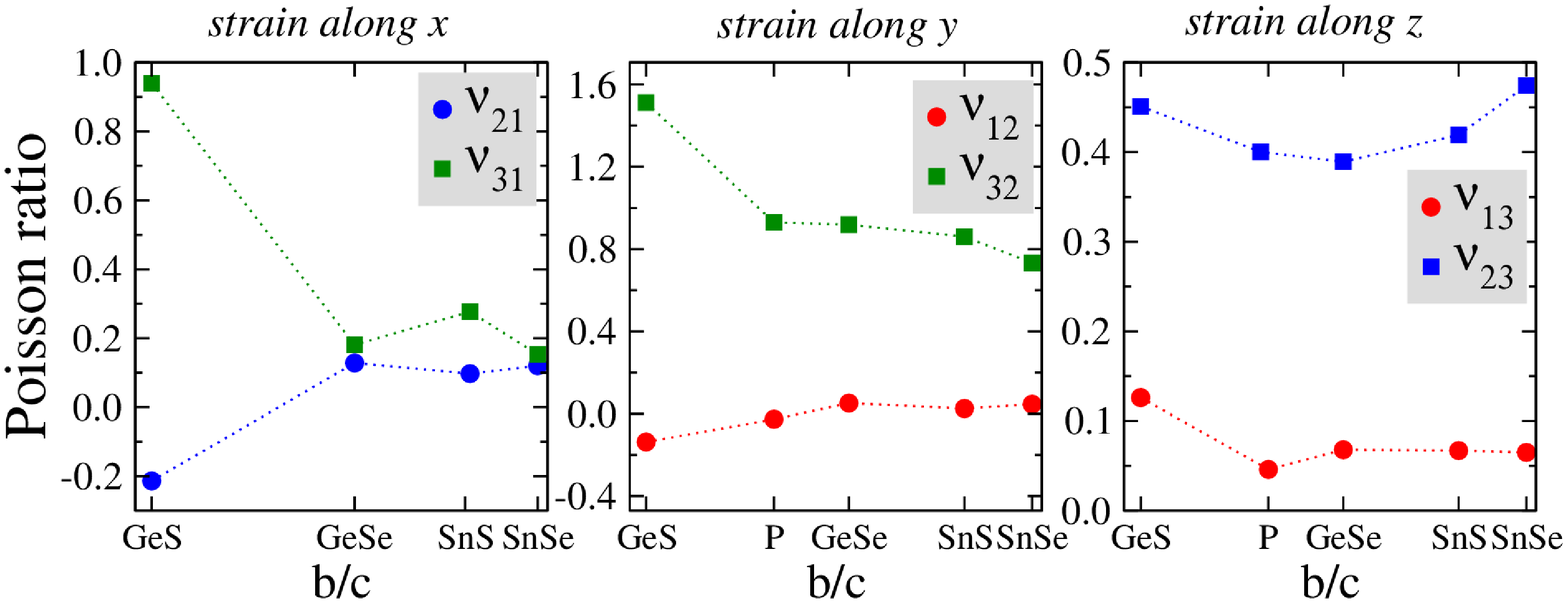}}
   \caption{\small (Color online) Poisson ratio versus the degree of anisotropy, given by the ratio 
                   between in-plane lattice constants $b$ and $c$, for monolayer 
                   crystals. The values for phosphorene (P) taken from Ref.~\onlinecite{Jiang2014} are also included.}
   \label{PoissonXb_c}
 \end{figure}

In contrast, the ratios $\nu_{23}$ and $\nu_{32}$ for in-plane uniaxial stress are in the ranges 0.3-0.5 and 0.7-1.6.
The Poisson ratio is intrinsically linked with the anisotropy.
Plotting the Poisson ratio against the anisotropy of the monolayer crystals ($b$/$c$), 
it becomes evident why GeS departs from the other materials (Fig.~\ref{PoissonXb_c}). 
Amongst all four materials, GeS has the smallest $b/c$ ratio, and is located far away from GeSe, SnS and SnSe. 
This behavior is more evident in the plot of $\nu_{21}$ and $\nu_{31}$ (left panel in Fig.~\ref{PoissonXb_c}), obtained by applying uniaxial strain in the out-of plane $x$ direction.

\begin{table*}[ht]
\centering
\renewcommand{\arraystretch}{2.0}
\begin{tabular}{lcccccccccccccccc}
       \hline \hline
&&\multicolumn{2}{c}{$\varepsilon_{1}$}&& \multicolumn{2}{c}{$\varepsilon_{2}$}&& \multicolumn{2}{c}{$\varepsilon_{3}$}& & & &  \\
 \cline{3-4} \cline{6-7} \cline{9-10}
&\phantom{abcde}&$\nu_{21}$&$\nu_{31}$&\phantom{abcd}&$\nu_{12}$&$\nu_{32}$&\phantom{abcd}&$\nu_{13}$&$\nu_{23}$&\phantom{abcd}& $b/c$  &\phantom{abcd}& $Y_1$& $Y_2$ & $Y_3$  \\  \hline
SnS   && 0.097    & 0.277   &&  0.025   &   0.861 && 0.067    & 0.419     && 0.97  && 32.4    & 15.1  & 8.7    \\
SnSe  && 0.120    & 0.153   &&  0.046   &   0.733 && 0.065    & 0.474     && 0.99  && 30.4    & 16.2  & 11.3   \\
GeS   &&-0.214    & 0.940   && -0.137   &   1.512 && 0.126    & 0.451     && 0.84  && 26.8    &  7.8  & 2.2    \\
GeSe  && 0.128    & 0.181   &&  0.051   &   0.919 && 0.068    & 0.389     && 0.94  && 30.2    & 15.9  & 7.9    \\  \hline
P     &&     -    &    -    && -0.027   &   0.930 && 0.046    & 0.400     && 0.88  && 44.0    & 166.0 &  -     \\  \hline \hline
\end{tabular}
\caption{\small Poisson ratio and Young modulus (in GPa) for the monolayers monochalcogenides. The data for monolayer phosphorene from Refs.~\onlinecite{Jiang2014} and ~\onlinecite{apl-25-104} are included for comparison. $b/c$ is the ratio between in-plane lattice parameters.}
\label{Poisson-num}
\end{table*}

\subsection{Piezoelectric tensor} 

Non-centrosymmetric crystals display a change of polarisation $P_{i}$ under mechanical stress,
\begin{equation}
 P_{i} = d_{ij} \sigma_{j}.
 \label{deff}
\end{equation}
Equation~\ref{deff} shows that application of a stress $\sigma_{j}$ along the $j$-direction of a piezoelectric 
material induces a change of polarisation of magnitude $P_{i}$ in its $i$-direction. 
$P_{i}$ is related to $\sigma_{j}$ 
by a piezoelectric tensor with components $d_{ij}$. This is known as the direct piezoelectric effect. In 
the same way, the converse piezoelectric effect occurs when a strain $\varepsilon$ is induced in a material under 
an external electric field. The converse effect can be written:

\begin{equation}
 \varepsilon_{j} = d_{ij} E_{i},
 \label{ceff}
\end{equation}
where $E_{i}$ is the component of the applied electric field in the $i$ direction. The piezoelectric coefficients $d_{ij}$ in Eqs.~\ref{deff} and \ref{ceff} are the same, and the proof of such equality is based on thermodynamical arguments~\cite{nye}. In addition, other piezoelectric coefficients can be defined, for example $e_{ij}$, that directly relates $E$ and $\sigma$ by:

\begin{equation}
 e_{ij} = - \dfrac{\partial \sigma_{j}}{\partial E_{i}} = \dfrac{\partial P_{i}}{\partial \varepsilon_{j}}.
 \label{eij}
\end{equation}

For a constant electric field, the piezoelectric tensors $d_{ij}$ and $e_{ij}$ are related via the 
elastic tensor $C_{ij}$ as:

\begin{equation}
 e_{ij} = \sum_{k=1}^{6}d_{ik} C_{kj}.
 \label{de}
\end{equation}

However, in the case of a 2D system it is more meaningful to define a 2D polarisation per unit area, $P^{\rm 2D}_i$.
Thus, we redefine $e^{\rm 2D}_{ij}$
\begin{equation}
 e^{\rm 2D}_{ij} =  \dfrac{\partial P^{\rm 2D}_{i}}{\partial \varepsilon_{j}}.
 \label{eij2D}
\end{equation}

The number of non-zero coefficients $e^{\rm 2D}_{ij}$  are completely defined by the symmetry of the system. 
The layered group-IV monochalcogenides are centrosymmetric in bulk (and even-numbered layers) and therefore are not piezoelectric. 
However, single layers  belong to the polar space group $Pmn2_1$. 
In this space group, there are five non-zero piezoelectric constants: $e^{\rm 2D}_{15}$, $e^{\rm 2D}_{24}$, $e^{\rm 2D}_{31}$, $e^{\rm 2D}_{32}$, and $e^{\rm 2D}_{33}$, where we use the Voigt's notation~\cite{nye} with 1, 2 and 3 corresponding to $x$, $y$ and $z$ directions, respectively.\footnote{According to Voigt's notation, the indices are contracted according to 11~\textrightarrow~1, 22~\textrightarrow~2, 
11~\textrightarrow~1, 33~\textrightarrow~3, 23(32)~\textrightarrow~4, 13(31)~\textrightarrow~5, 
12(21)~\textrightarrow~6, so that $e^{\rm 2D}_{16}$ = $e^{\rm 2D}_{112}$ = $e^{\rm 2D}_{121}$, $e^{\rm 2D}_{25}$ = $e^{\rm 2D}_{223}$ = $e^{\rm 2D}_{232}$, and so on.} As for the calculation of the elastic constants, we consider only in-plane strain components 
for computation of the $e^{\rm 2D}_{ij}$ constants, which limit our discussion to $e^{\rm 2D}_{32}$, $e^{\rm 2D}_{33}$ and $e^{\rm 2D}_{24}$. 
The calculated values are presented in Table~\ref{piezo}.

\begin{equation}
e^{\rm 2D}_{ij} =  \left[
\begin{array}{c c c c c c c}
      0         &      0      &     0       &      0       &    e^{\rm 2D}_{15} &      0     \\
      0         &      0      &     0       &  e^{\rm 2D}_{24} &      0         &      0     \\        
   e^{\rm 2D}_{31}  & e^{\rm 2D}_{32} & e^{\rm 2D}_{33} &      0       &      0         &      0     \\ 
\end{array}
          \right] 
\end{equation}



\begin{table}[ht]
\centering
\renewcommand{\arraystretch}{2.0}
\begin{tabular}{llcccccccc}
  \hline  \hline    
  &&\phantom{ab}&\multicolumn{3}{c}{clamped-ion} &\phantom{abc} & \multicolumn{3}{c}{relaxed-ion} \\
      \cline{4-6} \cline{8-10}
    &    &&$e^{\rm 2D}_{32}$&$e^{\rm 2D}_{33}$&$e^{\rm 2D}_{24}$&&$e^{\rm 2D}_{32}$&$e^{\rm 2D}_{33}$&$e^{\rm 2D}_{24}$ \\ \hline
SnS &  PBE    &&  -4.73      &  0.29       &   -4.39    &&   0.76      & 23.36       &    15.70     \\ 
    & vdW-TS  &&  -5.01      &  0.47       &   -4.75    &&   2.27      & 18.94       &    15.54     \\ \hline

SnSe&  PBE    &&  -4.89      &  0.53       &   -4.67    &&   4.42      & 24.18       &    28.17     \\       
    & vdW-TS  &&  -4.67      &  0.58       &   -4.29    &&   6.73      & 30.25       &    30.82     \\ \hline

GeS & PBE     &&  -6.69      & -1.25       &   -7.10    &&  -4.97      &  7.28       &     0.37     \\           
    & vdW-TS  &&  -6.89      & -0.81       &   -7.08    &&  -4.64      &  8.83       &     2.05     \\ \hline

GeSe&  PBE    &&  -7.16      & -0.26       &   -7.37    &&  -3.00      & 13.26       &     8.25     \\ 
    & vdW-TS  &&  -7.11      &  0.35       &   -7.20    &&  -1.48      & 16.95       &     12.48    \\ \hline

h-BN    &&&  -3.71      &  3.71       &     -      &&  -1.38      &  1.38       &      -       \\
MoS$_2$ &&&  -3.06      &  3.06       &     -      &&  -3.64      &  3.64       &      -       \\
MoTe$_2$&&&  -2.98      &  2.98       &     -      &&  -5.43      &  5.43       &      -       \\\hline \hline 
\end{tabular}
\caption{\small Nonzero ion-clamped and relaxed-ion piezoelectric coefficients (10$^{-10}$ C/m). 
                The tensor components are calculated according to $e_{ij}^{0} = - \frac{\partial \sigma_{i}}{\partial E_{j}}$, where the values for the other 2D materials h-BN, MoS$_2$ and MoTe$_2$ are taken from Ref.~\cite{jz3012436} and included for comparison.}
\label{piezo}
\end{table}

For the best of our knowledge, there is still no experimental data on the piezoelectric properties of 
odd-number layers group-IV monochalcogenides. However, we take for comparison single layer TMDCs and h-BN, 
of which piezoelectric coefficients have been theoretically calculated\cite{jz3012436}, 
and object of recent experimental measurements~\cite{piezo-exp}. 

Similar to  h-BN and TMDC,\cite{jz3012436} the piezoelectric elements $e_{32}$, $e_{33}$ and $e_{24}$
of group IV monochalcogenides are enhanced as we move downward in the periodic table.
The value of $e_{33}$ (for both relaxed and clamped-ion case)
seems also to be directly related to the degree of anisotropy of the materials as given by the
ratio between in-plane lattice parameters $b/c$.

For SnS and SnSe, $e_{33}$ is one order of magnitude higher than for other known 2D piezoelectric materials.
This element relates the polarisation along the $z$ direction, the polar direction of the crystal,
with the strain along the same direction.
Since, as we have discussed before, this structure is extremely ductile along the $z$ direction,
the corresponding piezoelectric response is very large.

Our results agree with those recently reported in Ref.~\onlinecite{apl-107-173104}, which also investigates piezoelectric properties of monochalcogenides. Although there are some small quantitative discrepancies, the overall qualitative results between both works are in quite good agreement.

It is also instructive to compare the piezoelectric coefficients of group-IV monochalcogenides with  
typical 3D piezoelectric materials. Employing once again the definition of an effective 
tickness $d_{0}$, we obtain $e_{33}$~=~4.11, 4.26, 1.36 and 2.37 C/m$^2$, for SnS, SnSe, GeS and GeSe, 
respectively. These coefficients are one to two orders of magnitude higher than the piezoelectric 
coefficients of the $\alpha$ quartz and four of its homeotypes $MXO_{4}$ ($M$ = Al, Ga, Fe; $X$=P,As)~\cite{PhysRevB.81.045107}.

\section{Conclusions}

The group-VI monochalcogenides SnS, SnSe, GeS and GeSe have an ortorhombic structure similar to phosphorene.
This structure results in in-plane anisotropy of the static permittivity, elastic constants and piezoelectric coefficients.
In this study, we investigated those properties both for monolayer and bulk systems by including van der Waals effects, highlighting the differences
resulting from the lower dimensionality.
The electric susceptibility in the 2D systems is known to have an $1/L$ dependence on the interlayer spacing.\cite{PhysRevB.84.085406}
This has been verified for electronic contribution to the low-frequency susceptibility.\cite{PhysRevB.88.045318}
In this work, we verified numerically that the same applies to the ionic contribution.
However, if we extrapolate the ionic contribution to the 2D permittivity to $L=a/2$,
where $a/2$ is the inter-layer spacing in bulk, we notice a great enhancement for SnS, SnSe and GeSe.
This is mainly due to the larger softening of the modes in the 2D systems, but also partially accounted for the effective Born charges in the 2D material.

In contrast, elastic constants remain nearly unchanged in monolayer, compared to bulk, if an equivalent volume of material is considered.
The most remarkable amongst the elastic properties of group-VI monochalcogenides is the Poisson ratio.
The Poisson ratio $\nu_{12}$ and $\nu_{21}$, relating strain and uniaxial strain
for the direction perpendicular to the plane and the armchair in-plane direction
are very small or even negative for GeS, the most anisotropic amongst the four materials.

Finally, while in bulk the presence of inversion symmetry forbids piezoelectricity,
in monolayer there is one polar direction coinciding with the $C_2$ axis ($z$).
Piezoelectric constants are higher than that recently predicted to occur in
 other 2D-systems, as hexagonal BN and transition metal dichalcogenide crystals.

\section*{Acknowledgements}
This work was supported by the National Research Foundation, Prime Minister Office, Singapore,
under its Medium Sized Centre Programme and CRP
award ``Novel 2D materials with tailored properties: beyond graphene" (Grant number R-144-000-295-281).
The first-principles calculations were carried out on the GRC high-performance computing facilities.

\label{Bibliography}
\bibliographystyle{unsrtnat} 
\bibliography{Bibliography} 

\end{document}